# A new Frequency Estimation Sketch for Data Streams

| 1st Author | 2nd Author | 3rd Author |
|---|---|---|
| 1st author's affiliation | 2nd author's affiliation | 3rd author's affiliation |
| 1st line of address | 1st line of address | 1st line of address |
| 2nd line of address | 2nd line of address | 2nd line of address |
| Telephone number, incl. country code | Telephone number, incl. country code | Telephone number, incl. country code |
| 1st author's email address | 2nd E-mail | 3rd E-mail |

## ABSTRACT
In this paper, we describe the formatting guidelines for VLDB Proceedings.

## 1. INTRODUCTION
### 1.1 Motivation
In data stream applications, one of the critical issues is to estimate the frequency of each item in the specific multiset. The multiset means that each item in this set can appear multiple times. The data streams in many applications are high-speed streams which contain massive data, such as real-time IP traffic, graph streams, web clicks and crawls, sensor database, and natural language processing (NLP) [2][6], etc. In these applications, the stream information needs to be recorded by the servers in real time. However, since the data streams in these applications are high-speed, the accurate recording and estimation of item frequencies is always impractical. An alternative approach for addressing this problem is to estimate the item frequencies based on probabilistic data structures, and this approach has been widely used in the high-speed data streams estimation [7][9]. Sketches is one of the typical probabilistic data structures, which are initially designed for the estimation of item frequencies in data streams [10][15]. At present, the sketches have been used in many different scenarios, such as sparse approximation in compressed sensing [16], natural language processing [17, 18], data graph [19, 20], and more [21]. In this paper, we mainly focus on the sketches used for frequency estimation.

### 1.2 Limitations of Previous Works
There are already many sketches have been proposed, such as the CM-sketch, CU-sketch, C-sketch, A-sketch, [][]. In these sketches, the counter size and the number counters in each layer are the same, which is shown in Fig. 1, so we call this kind of sketch as the rectangular-structure sketch, shorted as *r*-structure. The CM-sketch, CU-sketch, C-sketch, and A-sketch are all *r*-structure. Even the *r*-structure sketch has been widely used, it still has disadvantage.

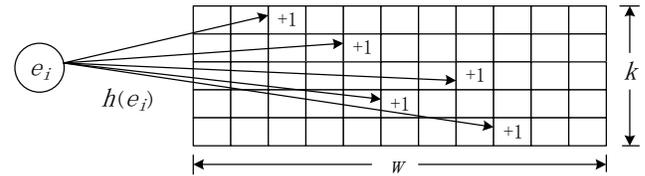

Fig. 1. The *r*-structure sketch

The main disadvantage of the *r*-structure sketch is that the ability of r-structure sketch is limited, i.e., the *r*-structure sketches cannot give consideration to both space usage, capacity of sketch (the capacity is defined as the maximum counter size of sketch), and estimation accuracy. This is because the frequencies of items are often highly skewed in real data streams [7, 9]. For many real data streams, such as the IP-trace streams and the real-life transactional dataset, most items in these data streams are cold (i.e., the frequency is low), while only a few items are hot (i.e., the frequency is high). So, for the *r*-structure sketches, it is difficult to find a proper counter size to fit these highly skewed data streams. For example, the frequencies of most items are cold in IP-trace stream (such as, smaller than 15), and the frequencies of only a few hot items are larger than 15,000. In this case, on one hand, if each counter is set with 4 bits, the space usage will be very small and the estimation accuracy of cold items is very high. However, the hot items will incur serious overflows in these counters, which can hardly be acceptable in many applications. Because, the hot items are always important in these applications. On the other hand, if we allocate 14 bits to each counter, the overflow of the hot items in the counters will be relieved and the estimation accuracy of the hot items is improved. However, the space usage will be extremely large, and most of the counters will be wasted since most items are cold item in the real data streams, i.e., the space utilization ratio of each counter is extremely low. This issue is hardly to be addressed by the existing *r*-structure sketches.

### 1.3 Proposed Approach and Advantages over Previous Works
For solving these issues, we propose the trapezoidal sketch in this paper, shorted as *t*-structure sketch. The t-structure sketch is shown in Fig. 2. As shown in Fig. 2, different with the r-structure sketch, in which the counter sizes in different layers are the same, the key innovation of *t*-structure sketch is that the counter sizes in different layers of sketch are different, and the counter sizes decrease in the *t*-structure sketch from top layer to down layer.

The proposed *t*-structure sketch has advantage over *r*-structure sketch. The main advantage of *t*-structure sketch is that its ability is better that that of r-structure sketch, i.e., it can give consideration to both space usage, capacity, and estimation accuracy. For instance, if we set the maximum counter size in t-structure sketch is the same as that in r-structure sketch, and the counter sizes in different layers are geometric series, then the capacity and the estimation accuracy of cold items in r-structure sketch and t-structure sketch are the same, the estimation accuracy of the hot items is slightly reduced; however, the space usage in t-structure sketch is much smaller than that in r-structure sketch. If the space usage in t-structure sketch and the r-structure sketch is the same and the counter sizes in different layers are geometric series, then the space usage is the same as that in r-structure sketch, the estimation accuracy of hot items is slightly lower than that in r-structure sketch; however, the capacity and the estimation accuracy of cold items is much higher than that in r-structure.

Since in the t-structure sketch, the estimation accuracy of the hot items is slightly lower than that in r-structure sketch, we propose the probabilistic based estimation error reduction approach in this paper. This approach can reduce the estimation error of both hot items and cold items of t-structure sketch.

## 2. RELATED WORKS

There are three different kinds of approaches that used for item frequency estimation in databased: sketches and bloom filter based approach.

Many sketches have been proposed, such as CM sketches [8], CU sketches [22], Count sketches [23], Augmented sketches [7], Sf-sketch [24], one memory access sketch [25], and Pyramid sketch [N1]. The most typically and widely used sketch is the CM-sketch, which is proposed in [8]. In CM-sketch, there are $k$ counter arrays and each counter array contain $w$ counters. As shown in Fig. 1, the counter sizes are same in these arrays. In the construct phase, the recording process for an element $e_i$ is to first map $e_i$ into $k$ different positions $h_1(e)\%w, h_2(e)\%w, \cdots, h_K(e)\%w$ by using $k$ distinct hash functions $h_1(\cdot), h_2(\cdot), \cdots, h_k(\cdot)$, where the symbol % represents modular operation and the output of these hash functions is within $[1, w]$. Initially, all the counters in trapezoidal cm-sketch is set to 0. When we increase the count of an element, all the corresponding $k$ counters in the different layers are increased by 1. When querying an item $e_i$, the CM sketch reports the minimal one among these k mapped counters. Different with the CM-sketch in which k mapped counters are all increased by 1 during the insert process, the CU sketch [22] only increases the smallest counter among the k mapped counters; so, the CU sketch also has the highest accuracy in these algorithms when using the same amount of memory. Similar approach can be found in [23], in which the Counter sketch is proposed. The main difference between CM sketch and Counter sketch is that in Counter sketch, there are two hash functions in each array. This approach can reduce the estimation error. The Augmented sketch is proposed in [7]. Different with the above-mentioned sketches, the A sketch improves the estimation accuracy by adding one filter to an existing sketch to dynamically capture hot items. The A sketch is accurate to the items which exist in its filter. However, the A sketch is complexity, and the query speed and the update are slow. More sketches can be found in [9], [24-N1]. Unfortunately, all these sketches are the r-structure sketch; so, they all suffer the advantages that mentioned in Section 1.2.

The bloom filter based solutions include Counting Bloom Filters [27], Spectral Bloom Filter (SBF) [28], Dynamic Count Filter (DCF) [29] and, Shifting Bloom filter [30, 31]. CBF is quite similar to the CM sketches; in CBF, only one array is used and k hash functions are hired in this array. The SBF, DCF, and Shifting bloom filter are all the improvements of CBF, which can be used to record the item frequency.

## 3. TRAPEZOIDAL SKETCH

In this section, we first describe the key innovation of *t*-structure sketch, including the construction phase and query phase. Then we propose two simple but high effective *t*-structure sketches, i.e., the space saving sketch and the capacity improvement sketch.

### 3.1 The basic *t*-structure sketch

The *t*-structure sketch is similar but different with the *r*-structure sketch, which is shown in Fig. 2. As shown in Fig. 2, the counter sizes in different layers are different in *t*-structure sketch; however, these are equal in *r*-structure sketch. The construction phase and query phase of *t*-structure sketch are the same as that in *r*-structure sketch.

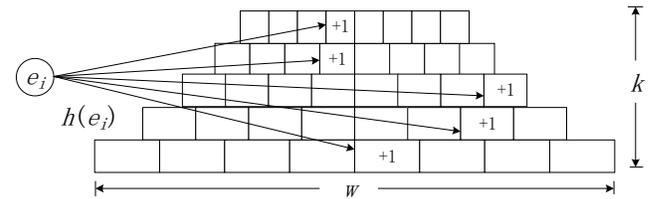

Fig. 2. The *t*-structure sketch

In construction phase, the *t*-structure sketch first constructs $k$ counter arrays. In each counter array, there are $w$ counters. The counter size in *ith* counter array (we call it *ith* layer in this paper) is $S_i$, where $1 \leq i \leq k$. Therefore, the sizes of counters in different layers in *t*-structure sketch are $\{S_1, S_2, ..., S_k\}$. In *t*-structure sketch, we assume that $S_i > S_j$ if and only if $i > j$ holds, i.e., the counter size increased from top layer to down layer. Then, the number of bits of the counter in *ith* layer can be calculated as $\log_2 S_i$. For instance, in Fig. 2, the counter size in the first layer is $S_1$ and the number of bits of the counter in this layer is $\log_2 S_1$.

Suppose the *t*-structure sketch needs to record *n* distinct items $e_1, e_2, \cdots, e_n$. Then, the recording process for item $e_i$ is to first map $e_i$ into $k$ different positions $h_1(e)\%w, h_2(e)\%w, \cdots, h_K(e)\%w$ by using $k$ distinct hash functions $h_1(\cdot), h_2(\cdot), \cdots, h_k(\cdot)$, where the symbol % represents modular operation and the outputs of these hash functions are within $[1, w]$. Initially, all the counters in *t*-structure sketch is set to 0. When we increase the count of an item, all the corresponding $k$ counters in different layers are increased by 1. The *t*-structure sketch supports deletion operation. If the deletion operation is needed, then all these $k$ related counters in different layers is reduced by 1. In this paper, we do not talk about the deletion operation.

In the query phase, the *t*-structure sketch obtains the frequency of items $e_i$ based on the following steps.

1. The sketch computes *k* hash functions $h_1(e)\%w, h_2(e)\%w, \cdots, h_k(e)\%w$ to determine *k* positions that the $e_i$ is mapped to in the *k* layers of *t*-structure sketch;

2. Reading the values of these *k* counters;

3. Chosen the minimum value in these $k$ counters as the estimated frequency of item $e_i$.

However, different with the $r$-structure sketch, the counter will overflow in $t$-structure sketch because the counter size is small in high layer (the high layer is the layer whose index number is large; for instance, in Fig.2, the $1st$ layer is high layer, the $kth$ layer is the low layer). For instance, as shown in Fig. 2, suppose the frequency of item $e_i$ is $f_{e_i}$ and $f_{e_i} \geq S_j$, then the $j$ counters from the first layer to the $jth$ layer that the $e_i$ mapped to will overflow. There are two situations can cause counter overflow in $t$-structure sketch. First, the frequency of item $e_i$ is larger than the counter size in $jth$ layer, i.e., $f_{e_i} > S_j$. Second, the frequency of item $e_i$ is smaller than the counter size in $jth$ layer ($f_{e_i} < S_j$). However, due to the hash collision, for the counter $x$ which the item $e_i$ is mapped to will be contributed by other items, i.e., the value of counter $x$ is not only the frequency of item $e_i$, it may also contain the frequencies of the other items. We call this as the estimation noise of item $e_i$, denoted as $X_i$. Therefore, due to the noise from other items, the counter overflows, i.e., $f_{e_i} + X_i > S_j$. For the first situation, the probability that the counters whose sizes smaller than $f_{e_i}$ will overflow is 1. However, the counter overflow is probabilistic in the second situation. We give the conclusion as follows.

**Corollary 1.** In $jth$ layer of the space saving $t$-structure sketch, the probability that the estimation noise of item $e_i$ cannot cause counter overflow in this layer is at least $1 - \frac{\|f_{-e_i}\|_1}{w(S_j - f_{e_i})}$, i.e., $Pr[X_i \leq S_j - f_{e_i}] \geq 1 - \frac{\|f_{-e_i}\|_1}{w(S_j - f_{e_i})}$.

Proof. In $jth$ layer, suppose the real frequency of item $e_i$ is $f_{e_i}$, which is smaller than $S_j$. The noise is $X_i$, then Markov's inequality, we have:

$$Pr[f_{e_i} + X_i \geq S_j] = Pr[X_i \geq S_j - f_{e_i}] \leq \frac{E(X_i)}{S_j - f_{e_i}} \quad (22)$$

In sketch, the $X_i$ can be calculated as:

$$X_i = \sum_{j \in [n] \setminus e_i} f_j Y_{i,j} \quad (11)$$

where $Y_{i,j}$ represents that the item $e_j$ is mapped to the counter that belong to item $e_i$. Moreover, we define

$$Y_{i,j} = \begin{cases} 1, & e_j \text{ is mapped to } e_i'\text{s counter} \\ 0, & e_j \text{ is not mapped to } e_i'\text{s counter} \end{cases} \quad (12)$$

Since the results of the hash functions are uniformly distributed, so the average of $Y_{i,j}$ can be calculated as:

$$E[Y_{i,j}] = \frac{1}{w} \quad (13)$$

Thus, the distribution of $X_i$ is $X_i \sim \left(f_j, \frac{1}{w}\right)$. Then, the average of $X_i$ is:

$$E[X_i] = \sum_{j \in [n] \setminus e_i} \frac{f_j}{w} = \frac{\|f\|_1 - f_{e_i}}{w} = \frac{\|f_{-e_i}\|_1}{w} \quad (14)$$

where $\|f_{-e_i}\|_1$ is frequency sum of the other items except $e_i$, $\|\cdot\|_1$ is the first-order norm and defined as: $\|a\|_1 = \sum_{i=1}^m |a_i|$. So, the probability that $X_i < S_j - f_{e_i}$ is:

$$Pr[X_i \leq S_j - f_{e_i}] = 1 - Pr[X_i \geq S_j - f_{e_i}]$$

$$\geq 1 - \frac{E(X_i)}{S_j - f_{e_i}} = 1 - \frac{\|f_{-e_i}\|_1}{w(S_j - f_{e_i})} \quad (23)$$

Thus, the Corollary 5 holds. ∎

During the querying process, the values in the counters which are overflow will be useless. Thus, only the minimum value in the layers whose counters are not overflow can be chosen as the estimated frequency of item $e_i$. For instance, as shown in Fig.2, suppose the frequency of element $e_i$ is $f_{e_i}$ and $f_{e_i} \geq S_j$, then the values of the counters from the first layer to the $jth$ layer that $e_i$ mapped to are useless; only the counters from the $(j + 1)th$ layer to the $kth$ layer can be used to estimate the frequency of item $e_i$. So, the same as the $r$-structure sketch, the real frequency of an item can never be larger than the estimated value in $t$-structure sketch.

In the following of this section, we will propose two simple $t$-structure sketches, which are the space saving $t$-structure sketch (in which the maximum counter size is the same as that in the $r$-structure sketch) and the capacity improvement $t$-structure sketch (in which the total space use is the same as that in the $r$-structure sketch). In these two $t$-structure sketches, for simplify, we assume the counter sizes in different layers are geometric series.

### 3.2 The space saving $t$-structure sketch

For saving space in $t$-structure sketch, let the capacity in space saving $t$-structure sketch is the same as that in $r$-structure sketch, i.e., the maximum counter size in space saving $t$-structure sketch is $B$, which is the same as that in $r$-structure sketch. Therefore, as shown in Fig. 2, the counter size in $kth$ layer of $t$-structure sketch is $B$. The counter size in $ith$ level is $S_i$ and the number of bits of this counter is $\log_2 S_i$. For simplify, we assume the counter sizes in different layer are geometric series. This means for the $ith$ layer, the counter size can be calculated as:

$$S_i = S_1 \cdot d^{(i-1)} \quad (1)$$

where $d$ is the common ratio in geometric series. Moreover, $S_k = S_1 \cdot d^{(k-1)} = B$. The construction phase and query phase are the same as the basic $t$-structure sketch, which is introduced in Section 3.1.

In the following of this section, we will analyze the properties of the space saving $t$-structure sketch. The properties include the saved space, the error probability, and the estimation error bound.

*Saved Space.* One of the main advantages of space saving $t$-structure sketch is that it can reduce the space usage greatly. For the saved space, we have the conclusions as follows.

**Corollary 1.** The space saving $t$-structure sketch can reduce the space usage greatly, and the reduction ratio is $\gamma = \frac{\log_2 d^{(k-1)}}{\log_2 B^2}$, where $d < B^{\frac{1}{k-1}}$.

Proof. In $r$-structure sketch, the bits of each counter are $\log_2 B$, the number of layers is $k$, and the number of counters is $w$ in each layer. In space saving $t$-structure sketch, the maximum bits of counter are $\log_2 B$, the number of layers is $k$, and the number of counters is $w$ in each layer, too. So, the total bits in $r$-structure sketch is:

$$T_r = \log_2 B^{kw} \quad (2)$$

Since the counter sizes in different layers are geometric series in space saving $t$-structure sketch, the bits of counter in different layer from large to small are $B, \frac{B}{d}, \cdots, \frac{B}{d^{(k-1)}}$. Therefore, the value of $B$ and

$k$ should meet the requirements that $\frac{B}{d^{(k-1)}} > 1$, i.e., $d < B^{\frac{1}{k-1}}$. The counter size in different layers are geometric series means that the bits of counter in each later are arithmetic progression. Thus, the bits of counter in $i$th layer of space saving $t$-structure sketch can be calculated as:

$$z_i = \log_2 \frac{B}{d^{(i-1)}} \quad (3)$$

Then the total bits of counters in space saving $t$-structure sketch is:

$$T_{sp-t} = kw \log_2 B - \frac{k(k-1)}{2} w \log_2 d$$

$$= \log_2 \frac{B^{kw}}{d^{\frac{k(k-1)}{2}w}} \quad (4)$$

So, the space reeducation can be calculated as:

$$\tilde{T} = T_r - T_{sp-t} = \log_2 d^{\frac{k(k-1)}{2}w} > 0 \quad (5)$$

Thus, the reduction ratio is

$$\gamma = \frac{\tilde{T}}{T_r} = \frac{\log_2 d^{\frac{k(k-1)}{2}w}}{\log_2 B^{kw}} = \frac{\log_2 d^{(k-1)}}{\log_2 B^2} \quad (6)$$

Therefore, the Corollary 1 holds. ∎

For the space saving $t$-structure sketch, the values of $B$ and $k$ are constant, which are the same as that in $r$-structure sketch. So, if we want to reduce the space usage, the value of $d$ needs to be increased under the constraint that $d < B^{\frac{1}{k-1}}$. On one hand, the larger value of $d$, the more memory space is saved; however, the estimation accuracy of the hot items reduces, which will be proved in the following of this section. On the other hand, if the value of $d$ is small, the space saved by the space saving $t$-structure sketch is small. But the estimation accuracy of the hot item is high. So, the value of $d$ can be set based on the statistics properties of the data streams. For instance, if hot items are important in the data streams, the $d$ should be large to keep the estimation accuracy; if the space usage is critical, the $d$ should be small to reduce the space usage as much as possible.

*Error Probability.* The same as the $r$-structure sketch, due to the hash collusion, the estimation value in space saving $t$-structure sketch may also be error (we call it the error probability). For the error probability, we have the conclusion as follow.

**Corollary 2.** The probability that the estimated value is error in space saving $t$-structure sketch is $\rho_{sp-i} = \left(1 - \left(1 - \frac{1}{w}\right)^{n-1}\right)^{k-i}$, where $i$ means that for the counters that element $e_i$ mapped to, the counters in the first $i$ layers are all overflow, i.e., from first layer to $i$th layer.

Proof. For the $r$-structure sketch, suppose that there are $n$ different items, the frequencies of these items are $f_1, f_2, \cdots, f_n$. If item $e_i$ chooses the counter $x$, then the probability that the other items also choose counter $x$ is $\frac{1}{w}$; so, the probability that there are no other items choose counter $x$ is $1 - \frac{1}{w}$. Since there are $n-1$ different items, if all these $n-1$ items are not mapped to counter $x$, the probability is:

$$\left(1 - \frac{1}{w}\right)^{n-1} \quad (7)$$

So, the probability that at least one of these $n-1$ items is mapped to counter $x$ can be calculated as:

$$1 - \left(1 - \frac{1}{w}\right)^{n-1} \quad (8)$$

In $r$-structure sketch, the minimum value of these $k$ counters is chosen as the estimated value. So, if the estimated value is correct, the only needed is that the minimum value in these $k$ values is correct. Since there are $k$ layers in $r$-structure sketch, so the probability that the estimated result is not correct can be calculated as:

$$\rho = \left(1 - \left(1 - \frac{1}{w}\right)^{n-1}\right)^k \quad (9)$$

The (9) means that when all these $k$ values are not correct, then the estimation is false.

For the space saving $t$-structure sketch, the calculation is similar to that of $r$-structure sketch. However, in space saving $t$-structure sketch, the hot items will overflow in the layers whose counter size is small, so the probability will be different with the $r$-structure sketch. Assuming that item $e_i$ overflows in $i$th layer, then the probability that the querying result is error in space saving $t$-structure sketch can be calculated as:

$$\rho_i = \left(1 - \left(1 - \frac{1}{w}\right)^{n-1}\right)^{k-i} \quad (10)$$

This is because the counters which overflow are useless for the estimation. Only the counters which does not overflow can be used to frequency estimation. Thus, the Corollary 2 holds. ∎

The Corollary 2 demonstrates that the error probability increases in space saving $t$-structure sketch due to the data overflow in high layer. Moreover, the error probability in space saving $t$-structure sketch is $\frac{1}{\left(1-\left(1-\frac{1}{w}\right)^{n-1}\right)^i}$ times larger than that in $r$-structure sketch.

Thus, the smaller $i$ is, the less increasing is. This means that for the cold items, in which the value of $i$ is small, the estimation accuracy is high. When $i = 0$, the estimation accuracy in space saving $t$-structure sketch is the same as that in $r$-structure sketch.

*Estimation Error Bound.* Except the error probability, we also investigate the estimation error bound of the space saving $t$-structure sketch. The conclusion is shown as follows.

**Corollary 3.** In the space saving $t$-structure sketch, the probability that the estimation error is smaller than $\beta \|f_{-e_i}\|_1$ is at least $1 - \frac{1}{(w\beta)^{k-i}}$, i.e., $Pr\left[\hat{f}_{e_i} - f_{e_i} \leq \beta \|f_{-e_i}\|_1\right] \geq 1 - \frac{1}{(w\beta)^{k-i}}$, where $\|f_{-e_i}\|_1$ is sum of the other items except $e_i$, $\hat{f}_{e_i}$ is the estimated frequency, $f_{e_i}$ is the real frequency, and $i$ means the counter overflow occurs in $i$th layer.

Proof. For the estimation error bound, we want to guarantee that the estimation error within a certain range (i.e., $\beta \|f_{-e_i}\|_1$) is larger than a certain probability ($\delta$), i.e., $Pr\left[X_i \leq \beta \|f_{-e_i}\|_1\right] \geq \delta$.

Based on the Markov's inequality, we have:

$$Pr\left[X_i \geq \beta \|f_{-e_i}\|_1\right] \leq \frac{E[X_i]}{\beta \|f_{-e_i}\|_1} \quad (15)$$

The $E[X_i]$ is calculated in (), so (15) equals to:

$$Pr\left[X_i \geq \beta \|f_{-e_i}\|_1\right] \leq \frac{\|f_{-e_i}\|_1}{w\beta \|f_{-e_i}\|_1} = \frac{1}{w\beta} \quad ()$$

Thus, for one certain layer, the probability that the estimation error within a certain range can be calculated as:

$$Pr\left[X_i \leq \beta\|f_{-e_i}\|_1\right] \geq 1 - \frac{1}{w\beta} \quad (16)$$

Since the estimation value is the minimum value of these $k$ counters, we have:

$$Pr\left[\hat{f}_{e_i} - f_{e_i} \geq \beta\|f_{-e_i}\|_1\right]$$
$$= Pr\left[\min\{X_1, X_2, \ldots, X_k\} \geq \beta\|f_{-e_i}\|_1\right]$$
$$= \prod_{i=1}^{k} Pr\left[X_i \geq \beta\|f_{-e_i}\|_1\right] \quad (17)$$

Considering that there are $k$ layers and $k$ independent hash functions in $r$-structure sketch, the (17) equals to:

$$Pr\left[\hat{f}_{e_i} - f_{e_i} \geq \beta\|f_{-e_i}\|_1\right]$$
$$= \prod_{i=1}^{k} Pr\left[X_i \geq \beta\|f_{-e_i}\|_1\right]$$
$$\leq \frac{1}{(w\beta)^k} \quad (17)$$

Thus, the probability that the estimation error is smaller than $\beta\|f_{-e_i}\|_1$ in $r$-structure sketch is:

$$Pr\left[\hat{f}_{e_i} - f_{e_i} \leq \beta\|f_{-e_i}\|_1\right]$$
$$= 1 - Pr\left[\min\{X_1, X_2, \ldots, X_k\} \geq \beta\|f_{-e_i}\|_1\right]$$
$$= 1 - \prod_{i=1}^{k} Pr\left[X_i \geq \beta\|f_{-e_i}\|_1\right]$$
$$\geq 1 - \frac{1}{(w\beta)^k} \quad (18)$$

For the space saving $t$-structure sketch, the calculation is similar but not same with that in $r$-structure sketch. In the space saving $t$-structure sketch, the counter may overflow in the high layer when the frequency of item is larger than the counter size. When the counter is overflow, we regard that the estimation error in this layer is 100%. Then the values in these layers will be ignored during the estimation. Only the values in the layers where the counters do not overflow can be used during the estimation. In the $t$-structure sketch, for the layers in which the counter does not overflow, the estimation accuracies of these layers are the same as that in $t$-structure sketch. For item $e_i$, suppose its frequency will overflow in $ith$ layer, then the probability that the estimation error is smaller than $\beta\|f_{-e_i}\|_1$ in space saving $t$-structure sketch is:

$$Pr\left[\hat{f}_{e_i} - f_{e_i} \leq \beta\|f_{-e_i}\|_1\right]$$
$$= 1 - Pr\left[\min\{X_{i+1}, X_{i+2}, \ldots, X_k\} \geq \beta\|f_{-e_i}\|_1\right]$$
$$= 1 - \prod_{i=1}^{k} Pr\left[X_i \geq \beta\|f_{-e_i}\|_1\right]$$
$$\geq 1 - \frac{1}{(w\beta)^{k-i}} \quad (19)$$

Thus, the Corollary 3 holds. ∎

From Corollary 3, we can conclude that the estimation error bound decreases in space saving $t$-structure sketch due to overflow in the high layer. Moreover, the estimation error bound in space saving $t$-structure sketch is $\frac{1-\frac{1}{(w\beta)^k}}{1-\frac{1}{(w\beta)^{k-i}}}$ times smaller than that in $r$-structure sketch. Thus, the smaller $i$ is, the less reducing is. This means that for the cold items, i.e., the value of $i$ is small, the estimation accuracy is high. When $i = 0$, the estimation error bound in space saving $t$-structure sketch is the same as that in $r$-structure sketch.

*Summary.* Based on the analysis above, we can conclude the properties of the space saving $t$-structure sketch. First, the space usage in space saving $t$-structure sketch is much less than that in $r$-structure sketch under the same capacity. Second, the same as $r$-structure sketch, there is no under-estimation of the items. Finally, by using the space saving $t$-structure sketch, the estimation accuracy of the cold items is the same as the $r$-structure sketch, while the estimation accuracy of the hot items is slightly lower than the $r$-structure sketch.

### 3.3 The capacity improvement $t$-structure sketch

In the capacity improvement trapezoidal cm-sketch, our purpose is to improve the capacity of the trapezoidal cm-sketch, i.e., the maximum counter size in the sketch. Thus, in the capacity improvement trapezoidal, let the space usage is similar to that in the traditional cm-sketch. Based on Corollary 1, the saved space in the space saving cm-sketch is $\log_2 d^{\frac{k(k-1)}{2}w}$, so the key innovation of the capacity improvement trapezoidal cm-sketch is that using the saved space to construct more higher layers with bigger counters over the $kth$ layer in space saving trapezoidal cm-sketch. The basic structure of capacity improvement trapezoidal cm-sketch is shown in Fig. 3. The construction phase and query phase are the same as the that introduced in Section 3.1.

As shown in Fig. 3, in the capacity improvement trapezoidal cm-sketch, the maximum counter size is larger than the traditional cm-sketch. Since capacity improvement trapezoidal cm-sketch utilizes the saved space by the trapezoidal structure to increase the number of layers, the first important issue is to calculate how many new layers can be constructed, which is investigated in Corollary 7.

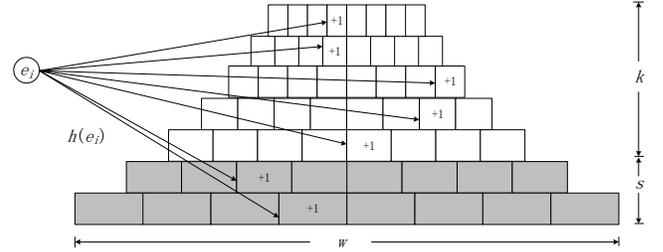

Fig. 3

**Corollary 7.** The maximum counter size of the capacity improvement trapezoidal cm-sketch is $c = d^{\tilde{s}}B$, where $\tilde{s} = \lfloor s \rfloor$ and

$$s = \frac{[(\log_2 dB^2)^2 + \log_2 d^{4k(k-1)}]^{\frac{1}{2}} - \log_2 dB^2}{\log_2 d^2}.$$

Proof. As shown in Fig. 3, the counter size in $kth$ layer is $B$, so the counter size in $(k+1)th$ layer is $dB$, the counter size in level $(k+2)th$ layer is $d^2B$, and so on. The space of the increased layers is equal to the saved space, and the counter bits in these increased layers are arithmetic progression, which are $\log_2 dB$, $\log_2 d^2B$, $\cdots$, $\log_2 d^sB$, the total bits of the increased counters can be calculated as:

$$Z = \log_2(dB)^{sw} + \log_2 d^{\frac{s(s-1)}{2}w} \quad (30)$$

She saved space by space saving structure is shown in (5), when $Z = \log_2 d^{\frac{k(k-1)}{2}w}$, we have:

$$\log_2(dB)^{sw} + \log_2 d^{\frac{s(s-1)}{2}w} = \log_2 d^{\frac{k(k-1)}{2}w} \quad (31)$$

Based on the Logarithmic algorithm, (31) equals to:
$$(dB)^{sw} \cdot d^{\frac{s(s-1)}{2}w} = d^{\frac{k(k-1)}{2}w} \quad (32)$$
which also equals to:
$$B^{2s} \cdot d^{s^2+s} = d^{k(k-1)} \quad (33)$$
Logarithm on the both sides of (33), we have:
$$\log_2 B^{2s} + \log_2 d^{s^2+s} = \log_2 d^{k(k-1)} \quad (34)$$
Based on the Logarithmic algorithm, the (34) equals to:
$$2s \log_2 B + (s^2 + s) \log_2 d = k(k-1)\log_2 d \quad (35)$$
Thus, $s$ can be calculated as:
$$s = \frac{[(\log_2 dB^2)^2 + \log_2 d^{4k(k-1)}]^{\frac{1}{2}} - \log_2 dB^2}{\log_2 d^2} \quad (36)$$

Since the value of $s$ that calculated by (36) may be not integer, the increased number of layers $\tilde{s}$ should be: $\tilde{s} = \lfloor s \rfloor$. Thus, the capacity of the capacity improvement cm-sketch, i.e., the counter size in $(k + \tilde{s})th$ layer, can be calculated as $c = d^{\tilde{s}}B$. ∎

As shown in (36), the value of $s$ is decided by both $k$ and $d$. Different values of $k$ and $d$ have different value of $s$. Moreover, the larger value of $s$, the more accurate of the estimation. Considering the constrain between $k$, $d$, and $B$ is $d < B^{\frac{1}{(k-1)}}$, our purpose can be expressed as:
$$\max \ s = \frac{[(\log_2 dB^2)^2 + \log_2 d^{4k(k-1)}]^{\frac{1}{2}} - \log_2 dB^2}{\log_2 d^2}$$
$$s.t. \quad d^{(k-1)} < B$$
$$d > 0 \quad (39)$$
$$k > 0$$

Moreover, the values of $B$, $d$, and $k$ are all integer.

In (39), the optimal value of $d$ and $k$ can be calculated when $s|'_d = 0$ and $s|'_K = 0$, respectively. For $k$, we have:
$$s|'_k = \frac{\partial}{\partial k} \left\{ \frac{[(\log_2 dB^2)^2 + \log_2 d^{4k(k-1)}]^{\frac{1}{2}} - \log_2 dB^2}{\log_2 d^2} \right\}$$
$$= \frac{(4k-2)}{\log_2 d^2} \cdot [(\log_2 dB^2)^2 + \log_2 d^{4k(k-1)}]^{-\frac{1}{2}} > 0 \quad (40)$$

Since $s|'_k > 0$ holds for all $k > 0$ and taking the constrain $d^{(k-1)} < B$ into account, the optimal value of $k$ when $d$ is constant can be calculated as $k_{max} = \lfloor \frac{\log_2 B}{\log_2 d} + 1 \rfloor$. When $d$ takes the minimum value as $d = 2$, the maximum optimal value of $k$ can be calculated as: $k_{max} = \lfloor \log_2 B + 1 \rfloor$. Since $k < \lfloor \log_2 B + 1 \rfloor$ and $B$ is even, $k_{max} = \log_2 B$. Then the value of $s$ under the maximum optimal value of $k$ can be calculated as:
$$s_1^* = \frac{[(\log_2 dB^2)^2 + \log_2 d^{4k(k-1)}]^{\frac{1}{2}} - \log_2 dB^2}{\log_2 d^2} \bigg|_{k=\log_2 B}$$
$$= \frac{[1+8(\log_2 B)^2]^{\frac{1}{2}} - \log_2 2B^2}{2}$$
$$\approx \log_2 B^{0.414} - \frac{1}{2} \quad (41)$$

For $d$, we have:
$$s|'_d = \frac{\partial}{\partial d} \left\{ \frac{[(\log_2 dB^2)^2 + \log_2 d^{4k(k-1)}]^{\frac{1}{2}} - \log_2 dB^2}{\log_2 d^2} \right\}$$

$$= \frac{\left\{ \frac{1}{2} f_d^{-\frac{1}{2}} \cdot \left[ \frac{4k(k-1)+2}{d \ln 2} \right] - \frac{1}{d \ln 2} \right\} \cdot \log_2 d^2 - \frac{f_d^{\frac{1}{2}}}{d \ln 2}}{(\log_2 d^2)^2} \quad (42)$$

where $f_d = (\log_2 dB^2)^2 + \log_2 d^{4k(k-1)}$. So, if $s|'_d = 0$, we have:
$$\left\{ \frac{1}{2} f_d^{-\frac{1}{2}} \cdot \left[ \frac{4k(k-1)+2}{d \ln 2} \right] - \frac{1}{d \ln 2} \right\} \cdot - \frac{f_d^{\frac{1}{2}}}{d \ln 2} = 0 \quad (43)$$

The (43) equals to:
$$\frac{1}{2} f_d^{-\frac{1}{2}} \cdot \left[ \frac{4k(k-1)+2}{d \ln 2} \right] = \frac{f_d^{\frac{1}{2}}}{d \ln 2} \cdot \frac{1}{\log_2 d^2} + \frac{1}{d \ln 2} \quad (44)$$

Since $d > 0$, so we have:
$$f_d^{-\frac{1}{2}} \cdot [2k(k-1) + 1] = \frac{f_d^{\frac{1}{2}}}{\log_2 d^2} + 1 \quad (45)$$
$$[2k(k-1) + 1] = \frac{f_d}{\log_2 d^2} + f_d^{\frac{1}{2}} \quad (46)$$

In (46), the analytic solutions of $d$ do not exist. Suppose the optimal value of $d$ is $d^*$, then $k = \lfloor \frac{\log_2 B}{\log_2 d^*} + 1 \rfloor$. Thus, the optimal value of $s$ under the optimal value of $d$ is:
$$s_2^* = \frac{[(\log_2 d^*B^2)^2 + \log_2 (d^*)^{4k(k-1)}]^{\frac{1}{2}} - \log_2 d^*B^2}{\log_2 (d^*)^2} \quad (47)$$

From (41) and (47), we conclude that:
$$s^* = max\{s_1^*, s_2^*\} \quad (48)$$

Therefore, the number of layers that can be increased in capacity improvement trapezoidal cm-sketch can be calculated as:
$$\tilde{s} = \lfloor s^* \rfloor = max\{\lfloor s_1^* \rfloor, \lfloor s_2^* \rfloor\} \quad (49)$$

Thus, there will be three principles on decide the value of $d$ and $k$.

1. If $\lfloor s_1^* \rfloor > \lfloor s_2^* \rfloor$ then $k^* = \log_2 B$ and $d^* = 2$.

2. If $\lfloor s_1^* \rfloor < \lfloor s_2^* \rfloor$, then $d = d^*$, $k = \lfloor \frac{\log_2 B}{\log_2 d^*} + 1 \rfloor$.

3. If $\lfloor s_1^* \rfloor = \lfloor s_2^* \rfloor$, then if $s_1^* - \lfloor s_1^* \rfloor > s_2^* - \lfloor s_2^* \rfloor$, $k^* = \log_2 B$ and $d^* = 2$; otherwise, if $s_1^* - \lfloor s_1^* \rfloor < s_2^* - \lfloor s_2^* \rfloor$, $d = d^*$, $k = \lfloor \frac{\log_2 B}{\log_2 d^*} + 1 \rfloor$.

The principle 1 and 2 are easy to be understood. Because the larger value of $s$, the higher capacity of the cm-sketch and the more accuracy of the frequency estimation. Therefore, in these two principles, we choose the solution which can make the $s$ get the maximum value as the final solution. The idea in principle 3 can be explained as follows. The value of $s_1^*$ and $s_2^*$ maybe not integer, so the value of $\lfloor s_1^* \rfloor$ and $\lfloor s_2^* \rfloor$ may equal (this is quite possible). When $\lfloor s_1^* \rfloor = \lfloor s_2^* \rfloor$, the maximum capacity and the accuracy are the same under these two optimal solutions. However, since $s_1^* \neq s_2^*$, the values of $s_1^* - \lfloor s_1^* \rfloor$ and $s_2^* - \lfloor s_2^* \rfloor$ are different. This value is the space that can be saved in the capacity improvement trapezoidal cm-sketch. Moreover, because $\lfloor s_1^* \rfloor = \lfloor s_2^* \rfloor$, the larger value of $s_1^*$ and $s_2^*$ makes the values of $s_1^* - \lfloor s_1^* \rfloor$ and $s_2^* - \lfloor s_2^* \rfloor$ larger. Thus, the larger value of $s_1^*$ and $s_2^*$ will be chosen as the final optimal solution.

In the following of this section, we will analyze the properties of the capacity improvement trapezoidal cm-sketch. The properties include the improved capacity, the saved space, the error probability, and the estimation error bound.

*Improved Capacity.* For the improved capacity, we have the conclusions as follows.

**Corollary 8.** The capacity of the capacity improvement trapezoidal cm-sketch is $d^{\tilde{s}}$ times larger than the traditional cm-sketch.

Proof. Based on Corollary 7, the maximum counter size in capacity improvement trapezoidal cm-sketch is $c = d^{\tilde{s}}B$, and the maximum counter size in traditional cm-sketch is $B$, so the capacity of the capacity improvement trapezoidal cm-sketch is $d^{\tilde{s}}$ times large than the traditional cm-sketch. Thus, the Corollary 8 holds. ∎

*Saved Space.* As discussed above, the capacity improvement trapezoidal cm-sketch not only can improve the capacity, it can also save the space usage (the saved space is the fractional part of $s_1^*$ and $s_2^*$ that calculated in (41) and (47)). The conclusion is shown in Corollary 9.

**Corollary 9.** For the capacity increasing trapezoidal cm-sketch, the saved space is $\theta_i = \log_2\left[B^{w(s_i^* - \lfloor s_i^* \rfloor)} \cdot d^{w(s_i^* - \lfloor s_i^* \rfloor)(s_i^* + 1)}\right]$, where $i = \{1,2\}$ represents the two optimal solutions.

Proof. Base on Definition 3, the fractional part of $s_i^*$ is $s_i^* - \lfloor s_i^* \rfloor$. Moreover, the counter size in $(s_i^* + 1)th$ layer is $B \cdot d^{(s_i^* + 1)}$, and the number of counters in $(s_i^* + 1)th$ layer is $w$. However, the residual bits (i.e., $s_i^* - \lfloor s_i^* \rfloor$) cannot construct a complete layer, so the number of counters that can be constructed by the residual bits is $w(s_i^* - \lfloor s_i^* \rfloor)$. Thus, the saved space in capacity improvement trapezoidal cm-sketch can be calculated as:

$$\theta_i = \log_2\left[B^{w(s_i^* - \lfloor s_i^* \rfloor)} \cdot d^{w(s_i^* - \lfloor s_i^* \rfloor)(s_i^* + 1)}\right] \quad (50)$$

Thus, the Corollary 9 holds. ∎

Based on the Corollary 9, the saved space in these three different principles can be calculated. Therefore, in the capacity improvement trapezoidal cm-sketch, not only the capacity is improved, but also the space is saved. However, this saved space is smaller than that in the space saving trapezoidal cm-sketch.

*Error Probability.* The calculation of error probability in the capacity improvement trapezoidal cm-sketch is similar with that of the space saving trapezoidal cm-sketch. However, due to the number of layers is different in these two structures, the values of error probabilities are different. Based on (10), the error probability of capacity improvement trapezoidal cm-sketch can be calculated as:

$$\rho_i = \left(1 - \left(1 - \frac{1}{w}\right)^{n-1}\right)^{k + \tilde{s} - i} \quad (56)$$

Based on (9) and (56), the error probability in capacity improvement trapezoidal cm-sketch is $\frac{1}{\left(1-\left(1-\frac{1}{w}\right)^{n-1}\right)^{i-\tilde{s}}}$ times larger than that in traditional cm-sketch, and $\left(1 - \left(1 - \frac{1}{w}\right)^{n-1}\right)^{\tilde{s}}$ times smaller than that in space saving trapezoidal cm-sketch. Thus, the smaller $i$ is, the less reduction is. This means that for the cold items, i.e., the value of $i$ is small, the estimation accuracy is high. When $i = 0$, the estimation accuracy in capacity improvement structure is better than that in traditional cm-sketch.

From (56), we can conclude that the error probability in capacity improvement cm-sketch may larger or smaller than that in traditional cm-sketch. This is decided by the value of $\tilde{s}$, i.e., if $\tilde{s} > i$, then $\frac{1}{\left(1-\left(1-\frac{1}{w}\right)^{n-1}\right)^{i-\tilde{s}}} < 1$, which means the error probability in capacity improvement cm-sketch is smaller than traditional cm-sketch; otherwise, if $\tilde{s} < i$, then $\frac{1}{\left(1-\left(1-\frac{1}{w}\right)^{n-1}\right)^{i-\tilde{s}}} > 1$, which means the error probability in capacity improvement cm-sketch is larger than traditional cm-sketch.

*Estimation Error Bound.* The estimation error bound of the capacity improvement trapezoidal cm-sketch is similar with that of the space saving trapezoidal cm-sketch. However, due to the number of layers is different in these two structures, the estimation error bounds are different. Based on (18), the estimation error bound of capacity improvement trapezoidal cm-sketch can be calculated as:

$$Pr\left[\hat{f}_{e_i} - f_{e_i} \leq \beta \|f_{-e_i}\|_1\right] \geq 1 - \frac{1}{(w\beta)^{k+\tilde{s}-i}} \quad (57)$$

Based on (16), (18), and (57), the estimation error bound in capacity improvement trapezoidal cm-sketch is $\frac{1 - \frac{1}{(w\beta)^k}}{1 - \frac{1}{(w\beta)^{k+\tilde{s}-i}}}$ times smaller than that in traditional cm-sketch, and $\frac{1 - \frac{1}{(w\beta)^{k-i}}}{1 - \frac{1}{(w\beta)^{k+\tilde{s}-i}}}$ times larger than that in space saving trapezoidal cm-sketch. Thus, the smaller $i$ is, the less reduction is. This means that for the cold items, i.e., the value of $i$ is small, the estimation accuracy is high. When $i = 0$, the estimation accuracy in capacity improvement structure is better than that in traditional cm-sketch.

From (56), we can conclude that the estimation error bound in capacity improvement cm-sketch may higher or lower than that in traditional cm-sketch. This is decided by the value of $\tilde{s}$, i.e., if $\tilde{s} > i$, then $\frac{1 - \frac{1}{(w\beta)^k}}{1 - \frac{1}{(w\beta)^{k+\tilde{s}-i}}} > 1$, which means the estimation error bound in capacity improvement cm-sketch is higher than traditional cm-sketch; otherwise, if $\tilde{s} < i$, then $\frac{1 - \frac{1}{(w\beta)^k}}{1 - \frac{1}{(w\beta)^{k+\tilde{s}-i}}} < 1$, which means the estimation error bound in capacity improvement cm-sketch is lower than traditional cm-sketch.

### 3.4 Potential Advantages of *t*-structure Sketch

The potential advantage of trapezoidal cm-sketch is its flexibility. By adjusting the counter size in each layer and the number of layers, many application requirements can be met, such as improving the capacity and estimation accuracy, reducing the space usage, etc. This is because the counter size and number of layers are critical to the performances of trapezoidal cm-sketch. Different counter size and number of layers have different performances on estimation accuracy, space usage, and capacity. For instance, as shown in Fig. 1, if the counter size in kth layer is the same as that in traditional cm-sketch, then the trapezoidal cm-sketch is space saving; if the space usage in Fig. 1 is the same as that in traditional cm-sketch, then the number of layers and the counter size in the maximum layer are larger than that in traditional cm-sketch. Moreover, if the statistical behavior of the data flows can be considered during setting the counter size and the number of layers, the performance can be improved further. For instance, in the IP-trace streams, in which the cold items are common, then most layers in sketch can be set with small counter size. By this, the space usage can be saved. Moreover, the saved space also can be used to construct more layers in sketch to improve the estimation accuracy. Therefore, if the machine learning technology can be introduced into the trapezoidal cm-sketch to learn the statistical behavior of the data flows and adjusting the counter size and number of layers in trapezoidal cm-sketch dynamically, the trapezoidal can benefit more. However, in this paper, we do not investigate this approach; this will be investigated in our further work.

## 4. Reducing the Estimation Error

As discussed above, the proposed trapezoidal cm-sketch is accurate for the cold items; the estimation error of the hot items is slightly higher than that of the traditional cm-sketch. Therefore, we propose the probabilistic based approach to reduce the estimation error of hot items.

In trapezoidal cm-sketch, suppose that the element $e_i$ is overflow in $ith$ layer, then the probability that the estimated value is error is $\rho_i$. In each counter, the average noise that contributed by other elements is $\frac{\|f_{-e_i}\|_1}{w}$, which is shown in (14). Thus, the average estimation error for the estimation value of $e_i$ can be calculated by:

$$\tilde{f}_{\varepsilon-min} = \rho_i \cdot \frac{\|f_{-e_i}\|_1}{w}$$
$$= \left(1-\left(1-\frac{1}{w}\right)^{n-1}\right)^{k-i} \cdot \frac{\|f_{-e_i}\|_1}{w} \quad (51)$$

Consequently, the estimation value in $ith$ layer which takes the probability of estimation error $\rho_i$ into account is:

$$\tilde{f}_{min} = f_{min} - \tilde{f}_{\varepsilon-min}$$
$$= f_{min} - \left(1-\left(1-\frac{1}{w}\right)^{n-1}\right)^{k-i} \cdot \frac{\|f_{-e_i}\|_1}{w} \quad (52)$$

where $v_i$ is the estimation value in $ith$ layer. Thus, the estimation value of the frequency of $e_i$ is $\tilde{v}_{e_i} = \min\{\tilde{v}_1, \tilde{v}_2, \ldots, \tilde{v}_k\}$ in space saving structure and $\tilde{v}_{e_i} = \min\{\tilde{v}_1, \tilde{v}_2, \ldots, \tilde{v}_{k+\tilde{s}}\}$ in capacity improvement structure. Note that based on the proposed probabilistic based approach, the estimation value may be smaller than the real frequency.

This approach is effective because the probability that the estimation is error is $\rho_i$. So, taking the $\rho_i$ into account during the estimation is reasonable and can reduce the estimation error. That means for the estimation approach shown in (52), the estimation is much more accurate than the traditional cm-sketch, which can be proved in the following.

In (16), the estimation accuracy of the traditional cm-sketch is calculated. Based on the approach proposed in (52), the noise from other items can be calculated as:

$$X_{pi} = \tilde{f}_{e_i} - \tilde{f}_\varepsilon - f_{e_i} = X_i - \tilde{f}_\varepsilon \quad (53)$$

However, in (53), when $\tilde{f}_{e_i} \neq f_{e_i}$, then $X_{pi} = X_i - \tilde{f}_\varepsilon$ holds; when $\tilde{f}_{e_i} = f_{e_i}$, then $X_{pi} = \tilde{f}_\varepsilon$. Thus, we have:

$$X_{pi} = \begin{cases} X_i - \tilde{f}_\varepsilon, & \text{Probability is } \rho_i \\ \tilde{f}_\varepsilon, & \text{Probability is } 1-\rho_i \end{cases} \quad (54)$$

Thus, we can calculate the average noise in the proposed approach is:

$$E[X_{pi}] = \rho_i(E[X_i] - E[\tilde{f}_\varepsilon]) + (1-\rho_i)E[\tilde{f}_\varepsilon] \quad (55)$$

Bring (51) into (55), we have:

$$E[X_{pi}] = \rho_i(1-\rho_i)E[X_i] + \rho_i(1-\rho_i)E[X_i]$$
$$= 2\rho_i(1-\rho_i)E[X_i] \quad (56)$$

So, the estimation accuracy of the trapezoidal cm-sketch can be calculated as:

$$Pr\left[X_{pi} \leq \beta\|f_{-e_i}\|_1\right] \geq 1 - \frac{E[X_{pi}]}{\beta\|f_{-e_i}\|_1}$$
$$= 1 - \frac{2\rho_i(1-\rho_i)E[X_i]}{\beta\|f_{-e_i}\|_1}$$
$$= 1 - \frac{2\rho_i(1-\rho_i)}{w\beta} \quad (57)$$

Therefore, the estimation accuracy in space saving structure is:

$$Pr\left[X_{pi} \leq \beta\|f_{-e_i}\|_1\right] \geq 1 - \frac{\left(1-\left(1-\frac{1}{w}\right)^{n-1}\right)^{k-i}}{w\beta} \quad (58)$$

The estimation accuracy in capacity improvement structure is:

$$Pr\left[X_{pi} \leq \beta\|f_{-e_i}\|_1\right] \geq 1 - \frac{\left(1-\left(1-\frac{1}{w}\right)^{n-1}\right)^{k+\tilde{s}-i}}{w\beta} \quad (59)$$

Based on the estimation accuracy calculated above, we have the conclusions as follows.

**Corollary 10.** The lower boundary of the estimation accuracy of trapezoidal cm-sketch by using the probabilistic based approach is $\varphi$ times higher than that of traditional cm-sketch, where $\varphi = \frac{w\beta-2\rho_i(1-\rho_i)}{w\beta-1}$.

Proof. The estimation accuracy of the trapezoidal cm-sketch and the traditional cm-sketch are shown in (16) and (57), respectively. Based on () and (), we have:

$$Pr\left[X_{pi} \leq \beta\|f_{-e_i}\|_1\right] - Pr\left[X_i \leq \beta\|f_{-e_i}\|_1\right] = \frac{1}{w\beta} - \frac{2\rho_i(1-\rho_i)}{w\beta}$$
$$= \frac{2\rho_i^2 - 2\rho_i + 1}{w\beta} \quad (60)$$

Let $f(\rho_i) = 2\rho_i(1-\rho_i)$, since $0 < \rho_i < 1$, $f(\rho_i) - 1 < 0$ holds. Thus, we have $Pr\left[X_{pi} \leq \beta\|f_{-e_i}\|_1\right] > Pr\left[X_i \leq \beta\|f_{-e_i}\|_1\right]$. Moreover, the ratio can be calculated as:

$$\varphi = \frac{1-\frac{2\rho_i(1-\rho_i)}{w\beta}}{1-\frac{1}{w\beta}} = \frac{w\beta-2\rho_i(1-\rho_i)}{w\beta-1} \quad (61)$$

This means that the Corollary 10 holds. ∎

Different with Corollary 10, based on (58) and (59), we find that the lower boundary of δ in capacity improvement cm-sketch is $\varphi$ times lower than that in space saving cm-sketch. Moreover, based on (61), the $\varphi$ can be calculated as:

$$\varphi = \frac{1-\frac{2\rho_1(1-\rho_1)}{w\beta}}{1-\frac{2\rho_2(1-\rho_2)}{w\beta}} = \frac{w\beta-2\rho_1(1-\rho_1)}{w\beta-2\rho_2(1-\rho_2)} \quad (62)$$

In (62), $f(\rho_i) = 2\rho_i(1-\rho_i)$ and $f(\rho_i) < 1$. Let $f(\rho_i)|'_{\rho_i} = 0$, then $\rho_i = \frac{1}{2}$. This means that when $\rho_i \in \left(0, \frac{1}{2}\right)$, $f(\rho_i)$ is an increasing function; when $\rho_i \in \left(\frac{1}{2}, 1\right)$, $f(\rho_i)$ is a decreasing function. So, the value of $\varphi$ can be decided as:

1. If both $\rho_1$ and $\rho_2$ are smaller than $\frac{1}{2}$, since $\rho_1 > \rho_2$, then $\varphi < 1$;

2. If both $\rho_1$ and $\rho_2$ are larger than $\frac{1}{2}$, since $\rho_1 > \rho_2$, then $\varphi > 1$;

3. If $\rho_1 > \frac{1}{2}$ and $\rho_2 < \frac{1}{2}$, when $\rho_1 + \rho_2 > 1$, then $\varphi > 1$; otherwise, when $\rho_1 + \rho_2 < 1$, then $\varphi < 1$.

**Corollary 11.** The upper boundary of noise in trapezoidal cm-sketch is $2\rho_i(1-\rho_i)$ times smaller than that in traditional cm-sketch.

Proof. Let the lower boundary of the estimation accuracy in trapezoidal cm-sketch and the traditional cm-sketch are the same, i.e., $Pr\left[X_{pi} \leq \beta_1 \|f_{-e_i}\|_1\right] = Pr\left[X_i \leq \beta \|f_{-e_i}\|_1\right]$, we have:

$$1 - \frac{2\rho_i(1-\rho_i)}{w\beta_1} = 1 - \frac{1}{w\beta} \quad (63)$$

The (63) equals to:

$$\beta_1 = 2\rho_i(1-\rho_i)\beta \quad (64)$$

Thus, the value of $\beta$ in trapezoidal cm-sketch is $2\rho_i(1-\rho_i)$ times smaller than that in traditional cm-sketch. This also means that the upper boundary of the noise in trapezoidal cm-sketch is $2\rho_i(1-\rho_i)$ smaller than that in traditional cm-sketch. ∎

Similar with the Corollary 11, according to (9), (10), and (57), the upper boundary of the noise in capacity improvement cm-sketch is $\mu$ times lower than that in space saving cm-sketch. Moreover, the $\mu$ can be calculated as:

$$\mu = \frac{2\rho_1(1-\rho_1)}{2\rho_2(1-\rho_3)} = \frac{1-\left(1-\left(1-\frac{1}{w}\right)^{n-1}\right)^{k-i}}{1-\left(1-\left(1-\frac{1}{w}\right)^{n-1}\right)^{k+\bar{s}-i}} \quad (65)$$

Similar to the analysis in (62), The value of $\mu$ can be decided as:

1. If both $\rho_1$ and $\rho_2$ are smaller than $\frac{1}{2}$, since $\rho_1 > \rho_2$, then $\mu > 1$;

2. If both $\rho_1$ and $\rho_2$ are larger than $\frac{1}{2}$, since $\rho_1 > \rho_2$, then $\mu < 1$;

3. If $\rho_1 > \frac{1}{2}$ and $\rho_2 < \frac{1}{2}$, when $\rho_1 + \rho_2 > 1$, then $\mu < 1$; otherwise, when $\rho_1 + \rho_2 < 1$, then $\mu > 1$.

## 5. PERFORMANCE EVALUATION

### 5.1 Experimental Setups

Datasets: We use three kinds of datasets as follows.

1) Real IP-Trace Streams: We obtain the real IP traces from the main gateway at our campus. The IP traces consist of ows, and each ow is identi_ed by its _ve-tuple: source IP address, destination IP address, source port, des-tination port, and protocol type. The estimation of item frequency corresponds to the estimation of number of packets in a ow. We divide 10M*10 packets into 10 datasets, and build a sketch with 1MB memory for each dataset. Each dataset includes around 1M ows. The ow characteristics in these datasets are similar. Take the _rst dataset as an example. The ow size ranges from 1 to 25,429 with a mean of 9.27 and a variance of 11,361. Note that 41.8% ows only have one packet. The length of items in each experimental dataset is 22 _ 46 bytes.

2) Real-Life Transactional Dataset: We downloaded this dataset from the website [34]. This dataset is built from a spidered collection of web html documents. It generates from each document a distinct transaction containing the set of all the distinct terms (items) appearing within the document itself. Since this raw dataset is very large, we use the first 8M items and estimate the frequency of each distinct items. For this transactional dataset with 8M items, it contains 352739 distinct items. The frequency ranges from 1 to 63724, with a mean of 21.35.

3) Synthetic Datasets: We generate 11 stream datasets following the Zipf [35] distribution with the _xed total frequency of items (10M) but di_erent skewnesses (from 0.0 to 1.0 with a step of 0.1) and di_erent numbers of distinct items. The larger the skewness is, the more skewed this dataset is. The main part of our generator's source codes come from a performance testing tool named Web Polygraph [36]. The maximum frequency of items in each dataset is 28 _ 21423. The length of items in each dataset is all 13 bytes.

Implementation: We have implemented the sketches of CM [8], CU [22], C [23] and A [7] in C++. We apply our Pyramid framework to these sketches, and the results are denoted as PCM, PCU, PC, and PA. For PCM, PC and PA, we apply the proposed techniques of counter-pair sharing and word acceleration (including word constraint, word sharing, and one hashing). For the PCU sketch, we additionally apply the Ostrich policy, which is only suitable for CU. The hash functions used in the sketches are implemented from the 32-bit or 64-bit Bob Hash (obtained from the open source website [37]) with di_erent initial seeds. For the sketches of CM and CU, we set the number of arrays to 4 and use 4 32-bit Bob Hashes. For the C sketch, we set the number of arrays to 4 and use 8 32-bit Bob Hashes4. For the A sketch, we set its _lter size to 32 items as the original paper [7] recommends. For the CM sketch contained in the A sketch, we set the number of arrays to 4 and use 4 32-bit Bob Hashes. For all the above four sketches, we set the counter size to 16 bits in the experiments with the IP traces and synthetic datasets, and to 24 bits in the experiments with the transactional dataset, so as to accommodate the maximal frequency of items. For the sketches of PCM, PCU, PC and PA, we set the number d of mapped counters to 4, the counter size _ to 4 bits, and the machine word size W to 64 bits. For each of these Pyramid sketches, we use 1 64-bit Bob hashes. In all our experiments with di_erent sketches, we allocate the same amount of memory, 1 MB, by default unless speci_ed otherwise. We allow the A sketch to use a small amount of additional memory for its _lter, which is _ 0:4KB and negligible when comparing with the total memory. Hence, the real memory allocation for A is 1MB + 0:4KB. The number of counters in each experiment can be easily calculated through the allocated memory size and the counter size in the sketch under use.

### 5.2 Metrics

*Average Absolute Error (AAE):* AAE is dened as where fi is the real frequency of item ei, bfi is the estimated frequency, and the is the query set.

*Average Relative Error (ARE):* ARE is defined as where the meaning of each notation is the same as that in AAE.

*Space efficient:* For evaluating the space efficient of different sketches, we evaluate two parameters: the absolute space usage and the space occupation ratio. The absolute space usage is the sizes of sketches. The space occupation ratio is the ratio of the bits that equal to 1 to the whole size of sketch.

*Capacity:* The capacity of sketch is the maximum counter size in the sketch; in other words, the capacity is the maximum frequency of item that can be recorded in sketch. We measure both the maximum counter size and the maximum frequency of item that can be recorded in sketch.

## 5.3 Accuracy

In this section, we evaluate the accuracy (include the AAE and ARE) of different sketches; moreover, the proposed probabilistic based estimation error reduction algorithm has been applied into these two sketches. The results of AAEs are shown in Fig. 3, Fig. 4, and Fig. 5, and the results of AREs are shown in Fig. 6, Fig. 7, and Fig. 8, respectively.

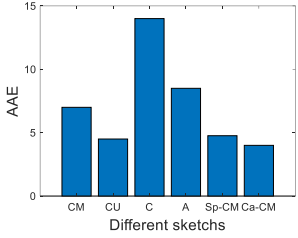

Fig. 3

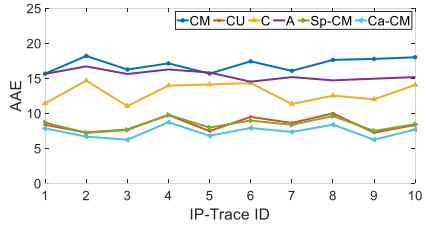

Fig. 4

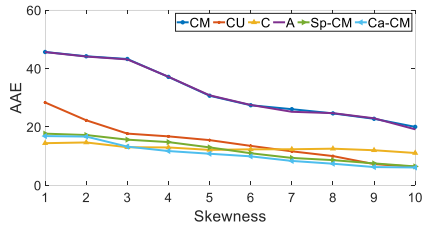

Fig. 5

*Our results shown that on different transactional dataset, the AAEs of CM, CU, C, and A sketch are 1.47, 0.95, 2.94, 1.79 times larger than that of Sp-CM sketch, and 1.85, 1.125, 3.5, and 2.125 times larger than that of Ca-CM.* Fig. 3 plots the AAEs of different sketches based on the transactional dataset. Moreover, in Fig. 3, the AAE of Sp-CM sketch is also larger than that in Ca-CM sketch. The reason why the AAEs of Sp-CM and Ca-CM sketch are better than that in CM sketch is because we apply the probabilistic based error reduction approach into the estimation, which reduce the estimation error. The reason why the AAEs of CM and A sketch are similar has been explained in [].

*Our results show that on different IP traces, the AAEs of CM, CU, C, and A sketch are 2.43, 1.34, 1.96, and 2.18 times larger than that of Sp-CM sketch, and 2.92, 1.6, 2.35, and 2.61 times larger than that of Ca-CM sketch.* Fig. 4 plots the AAEs of different sketches under different IP traces. Also, in Fig. 4, the AAE of Sp-CM sketch is larger than that in Ca-CM sketch. In each IP trace, the results are similar to that in Fig. 3. However, due to the difference between different data sets, the values of the results are different.

*Our results show that the on different skewness, the AAEs of CM, CU, C, and A sketch are 2.58, 1.6, 0.81, and 2.58 times larger than that of Sp-CM sketch, and 2.71, 1.69, 0.86, and 2.71 times larger than that of Ca-CM sketch.* Fig. 5 plots the AAEs of different sketches under different skewness, i.e., from 0 to 1. In Fig. 5, the AAE of Sp-CM sketch is larger than that in Ca-CM sketch. For different skewness, the smaller skewness is, the higher AAE is.

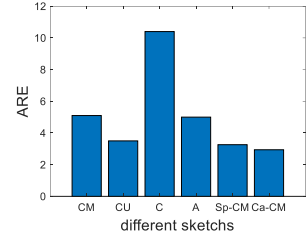

Fig. 6

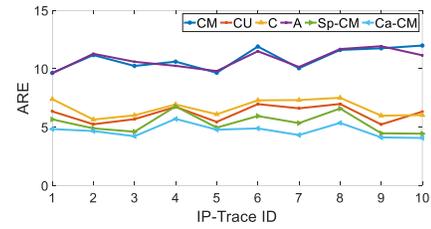

Fig. 7

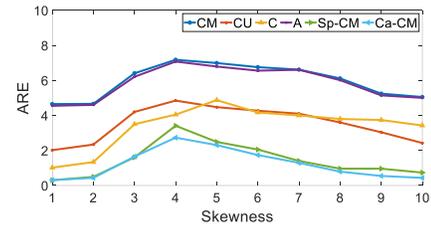

Fig. 8

*Our results shown that on different transactional dataset, the AAEs of CM, CU, C, and A sketch are 1.47, 0.95, 2.94, 1.79 times larger than that of Sp-CM sketch, and 1.85, 1.125, 3.5, and 2.125 times larger than that of Ca-CM.* Fig. 3 plots the AAEs of different sketches based on the transactional dataset. Moreover, in Fig. 3, the AAE of Sp-CM sketch is also larger than that in Ca-CM sketch. The reason why the AAEs of Sp-CM and Ca-CM sketch are better than that in CM sketch is because we apply the probabilistic based error reduction approach into the estimation, which reduce the estimation error. The reason why the AAEs of CM and A sketch are similar has been explained in [].

*Our results show that on different IP traces, the AAEs of CM, CU, C, and A sketch are 2.43, 1.34, 1.96, and 2.18 times larger than that of Sp-CM sketch, and 2.92, 1.6, 2.35, and 2.61 times larger than that of Ca-CM sketch.* Fig. 4 plots the AAEs of different sketches under different IP traces. Also, in Fig. 4, the AAE of Sp-CM sketch is larger than that in Ca-CM sketch. In each IP trace, the results are similar to that in Fig. 3. However, due to the difference between different data sets, the values of the results are different.

*Our results show that the on different skewness, the AAEs of CM, CU, C, and A sketch are 2.58, 1.6, 0.81, and 2.58 times larger than that of Sp-CM sketch, and 2.71, 1.69, 0.86, and 2.71 times larger than that of Ca-CM sketch.* Fig. 5 plots the AAEs of different

sketches under different skewness, i.e., from 0 to 1. In Fig. 5, the AAE of Sp-CM sketch is larger than that in Ca-CM sketch. For different skewness, the smaller skewness is, the higher AAE is.

The AAEs and AREs of Sp-CM and Ca-CM sketch are better than others because: 1) most items in the data set are cold items; 2) the application of the probabilistic based error reduction approach.

### 5.4 Space Usage Efficient
In this section, we evaluate the space usage efficient of different sketches. The results are shown in Fig. 9 and Fig. 10.

*Our results show that the space usage of CM, CU, C, and A sketch is 1.62 times and 1.12 times larger than that of Sp-CM and Ca-CM sketch, respectively*. Fig. 9 plots the absolute space usage of different sketches. Moreover, the space usage of Sp-CM sketch is smaller than that of Ca-CM sketch. The reason why the space usage in CM, CU, C, and A sketch is the same is because we set the sizes of these four sketches are equal to 1MB.

*Our results shown that the space usage ratio of Sp-CM sketch is about 1.71 times larger than that of CM, CU, C, and A sketch, and the space usage ratio of Ca-CM sketch is 1.64 times larger than that of CM, CU, C, and A sketch.* Fig. 10 plots the space usage ratios of different sketches. Moreover, the space usage ratio of Sp-CM sketch is larger than that of Ca-CM sketch.

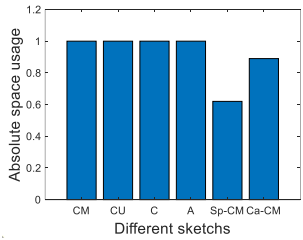

Fig. 9

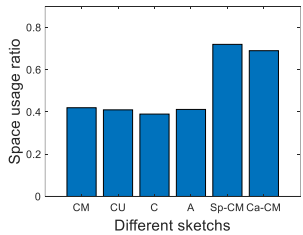

Fig. 10

### 5.5 Capacity
In this section, we evaluate the capacities of different sketches. The result is shown in Fig. 11.

*Our results show that the capacity of Ca-CM sketch is 4 times larger than that of CM, CU, C, A, and Sp-CM sketch.* Fig. 11 plots the capacities of different sketches. The reason why the capacities in of CM, CU, C, A, and Sp-CM sketch are the same is because we set the maximum counter size in these sketches is the same. Fig. 11 demonstrates the capacity of Ca-CM sketch on improving the maximum counter size in sketch.

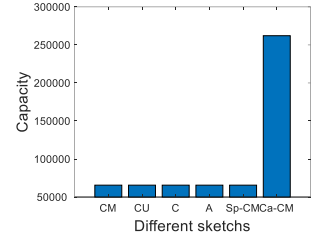

Fig. 11

## 6. CONCLUSIONS
The heading of a section should be in Times New Roman 12-point bold in all-capitals flush left with an additional 6-points of white space above the section head. Sections and subsequent sub-sections should be numbered and flush left. For a section head and a subsection head together (such as Section 3 and subsection 3.1), use no additional space above the subsection head.

## 7. ACKNOWLEDGMENTS
Our thanks to ACM SIGCHI for allowing us to modify templates they had developed.